\def\year{2021}\relax
\documentclass[letterpaper]{article} 
\usepackage{aaai21}  
\usepackage{times}  
\usepackage{helvet} 
\usepackage{courier}  
\usepackage[hyphens]{url}  
\usepackage{graphicx} 
\usepackage{natbib}  
\usepackage{caption} 
\usepackage{graphicx}
\usepackage{subcaption}
\usepackage{multirow}
\usepackage{amssymb,amsmath}   
\usepackage{diagbox}
\def\etal{et~al.}			  

\title{Exploiting Audio-Visual Consistency with\\
Partial Supervision for Spatial Audio Generation }
\author{
    Yan-Bo Lin \textsuperscript{\rm 1} and  Yu-Chiang Frank Wang\textsuperscript{\rm 1,2} \\

}
\affiliations{
    \textsuperscript{\rm 1}Graduate Inst. Communication Engineering, National Taiwan University, Taiwan\\
    \textsuperscript{\rm 2}ASUS Intelligent Cloud Services, Taiwan\\
    \{yblin98,ycwang\}@ntu.edu.tw
    
}

\begin{document}

\maketitle
\begin{abstract}
Human perceives rich auditory experience with distinct sound heard by ears. Videos recorded with binaural audio particular simulate how human receives ambient sound. However, a large number of videos are with monaural audio only, which would degrade the user experience due to the lack of ambient information. To address this issue, we propose an audio spatialization framework to convert a monaural video into a binaural one exploiting the relationship across audio and visual components. By preserving the left-right consistency in both audio and visual modalities, our learning strategy can be viewed as a self-supervised learning technique, and alleviates the dependency on a large amount of video data with ground truth binaural audio data during training. Experiments on benchmark datasets confirm the effectiveness of our proposed framework in both semi-supervised and fully supervised scenarios, with ablation studies and visualization further support the use of our model for audio spatialization.
\end{abstract}

\section{Introduction}\label{sec:intro}

\begin{figure*}[t!]
    \centering
	\includegraphics[width=0.9\linewidth]{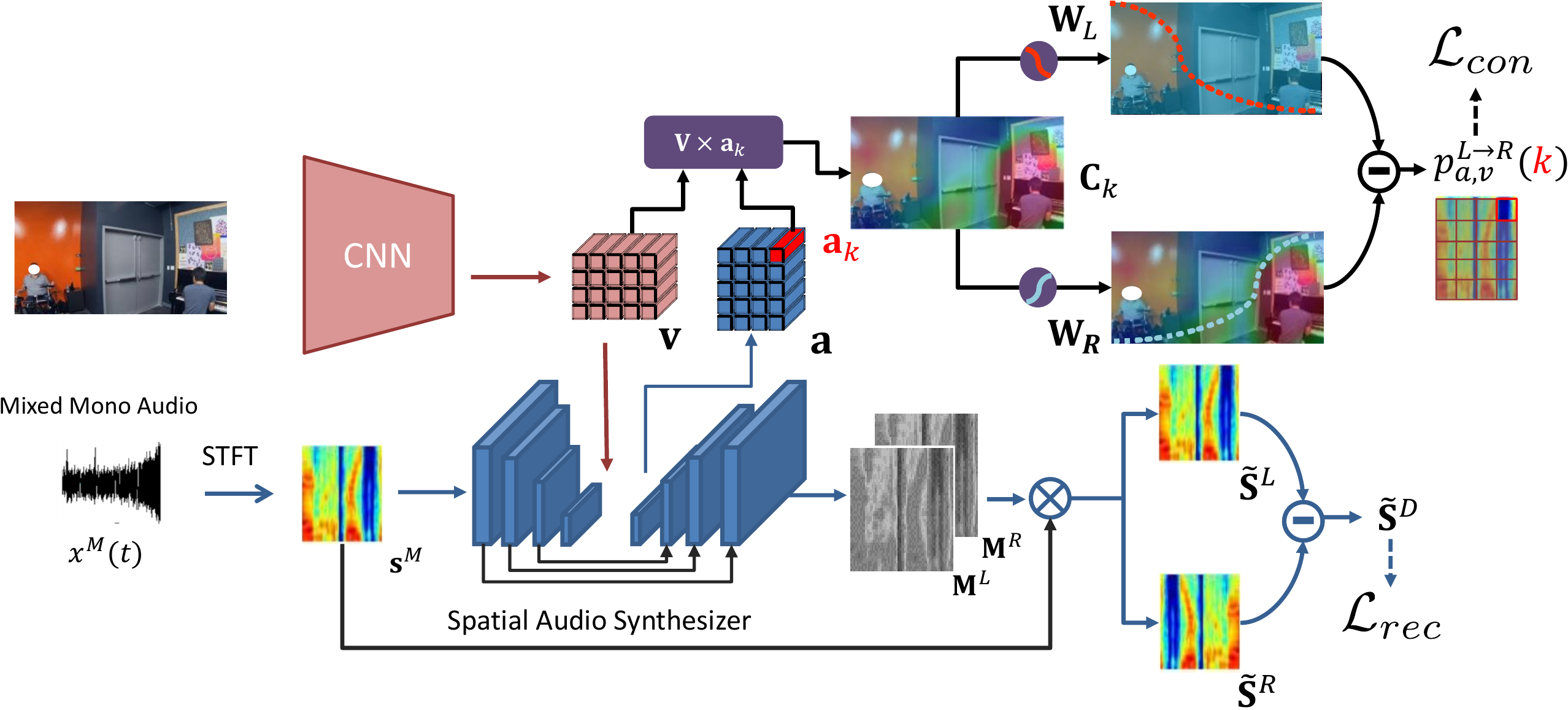}
     \caption{Overview of our proposed framework for binaural audio prediction. Our framework is composed of two main components, a spatial audio synthesizer taking monaural audio as inputs and predicting binaural audio outputs, and a CNN to extract visual features from the input video. In addition to jointly taking visual features into the generation of binaural audio outputs, we particularly observe audio-visual co-attention to identify spatial regions associated with audio components, with the observed consistency guides the learning process.}
	\label{fig:method}
\end{figure*}

Human beings are able to localize objects based on the sound heard by ears. The reason behind this ability is that, by parsing the audio difference between the two ears, human beings are able to infer spatial information of the sound origins (e.g., the drum is at the left hand side, and the piano is at the right hand side in Fig.~\ref{fig:method}), even if the audiences do not see and are not physically present in the scene. Thus, if one can design machines to measure the difference between the arrival times, including that between the frequency distributions perceive by left and right audio sensors, such machines would be able to perform sound localization accordingly.

However, most of the videos on social media contain only monaural audio signals (i.e., the same audio source heard by ears), which implicitly loses the spatial information of interest. Without the difference between two ears while perceiving sound, it is difficult for the users to immerse surroundings as if they were in the scenes. Therefore, the lack of spatial-related audio signals in media contents thereby diminishes the watching experiences of the users. 
To alleviate this issue, it would be desirable if one can convert monaural audio data into binaural ones. This is considered as the task of \textit{audio spatialization}~\cite{25d,25dclass,360gen,vr360,IEEEVR}, which is among active research topics in computer vision and signal processing, and with applications ranging from augmented reality (AR)~\cite{IEEEVR}, virtual reality (VR)~\cite{vr360}, social video sharing~\cite{25d,25dclass}, and audio-visual video understanding~\cite{AVSDN,eccv18_avel,Lin_2020_ACCV,tian2020unified}. Nevertheless, by observing visual data, generating audio outputs for left and right channels from a monaural audio input is a challenging task.

Since the visual content explicitly preserves the location of the sounding objects in a scene, it would be desirable to take videos accompanying monaural audio for recovering spatial sound information, i.e., to lift a flat audio signal into left-right spatial audio outputs. A number of methods~\cite{25d,25dclass,360gen} jointly considering spatial-audio features have been proposed. These methods are able to generate spatial audio signals associated with proper position of the sounding objects. With videos recording in binaural settings~\cite{25d,25dclass}, these methods simulate how monaural audio is presented on the mainstream media by mixing two channels audio into one channel. Thus, the resulting models of~\cite{25d,25dclass} are end-to-end trainable to perform audio spatialization with stimulated monaural audio inputs. Since the ground truth binaural outputs are available, the predicted binaural audio signals can be properly guided by the real binaural ones during training. While promising performance has been presented, collecting a sufficient amount of binaural audio datasets would be expensive. Moreover, the method in \cite{25dclass} requires additional scene classifiers during training, which might limit its generalization to videos with unseen content/scenes.


To overcome the above limitations, we propose a novel deep learning network for audio spatialization. By exploiting the visual cues across video frames, our model recovers binaural audio outputs from the input video with monaural recording. More specifically, we propose to identify audio-visual cross-modal correlation, which allows us to identify audio channels with the associated visual components. Such spatial information would guide the prediction of left and right-channel audio outputs throughout the video. Thus, the proposed framework would alleviate the dependency on the large amount of ground truth binaural videos required during training. In other words, our proposed framework can be realized in both supervised and semi-supervised settings. In our experiments, we extensively evaluate our proposed method on a benchmark dataset. From both qualitative and quantitative evaluation, our approach is shown to perform favorably against the state-of-the-art approaches in audio spatialization.

The contributions of this paper are highlighted below:
\begin{itemize}
  
  \item To recover binaural audio outputs from a video with only monaural audio recording, we exploit its audio-visual correlation to identify the sounding regions of interest in a scene without the associated visual ground truth information. This guides the audio recovery process with improved performances.
  
  \item The audio-visual correlation is calculated between the spectrogram of monaural audio and the visual features extracted across frames. This infers left-right audio-visual feature consistency, and can be viewed as a self-supervised learning strategy.

  \item Experiments on benchmark datasets demonstrate that our proposed module performs favorably against state-of-the-art approaches, and confirm that the learning scheme can be deployed in fully supervised and semi-supervised settings.
\end{itemize}

\section{Related Work}
\textbf{Audio-Visual Source Separation.} 
Mutual relationships between audio and visual data are exploited in the context of audio-visual source separation~\cite{nips00_av_sep}. Deep neural networks have been shown to be effective in utilizing visual cues for audio source separation~\cite{cocktail,av_eccv18_Owens}, musical instruments~\cite{pix,25d,sofm_iccv19,plus_iccv19,co_sep_iccv19} and objects~\cite{av_eccv18_sep}. Most of these methods adopt a ``mix and separate'' training strategy where the training videos are first mixed and separated afterward. For instance, MP-net~\cite{plus_iccv19} considers the sounds with larger energy which are first separated under all mixed sounds, and thus is removed from the mixture. As a result, sounds will smaller energies would keep emerging. In addition, the mixtures composed of any arbitrary number of sounds can also be separated by MP-Net.

To achieve object-level audio-visual source separation, Gao~\etal~\cite{co_sep_iccv19} propose a framework to bridge the localized object regions in a video with the corresponding sounds. The detected sounding objects can then be used to guide the learning process using unlabeled video data. Methods like~\cite{sofm_iccv19, sep-gesture-cvpr2020} utilize visual motions or body gesture to separate sound signals, and thus audio-visual source separation can be performed for different instruments. However, we note that source separation is different from the task of audio-spatial spatialization (as discussed later). The former identifies audio signals corresponding to a particular sounding objects of interest, while the latter needs to recover audio channel (e.g., left and right audio channels in binaural audio setting) which still includes audio signals from multiple sounding sources.

\textbf{Audio Generation from Visual Cues.} 
Recent works~\cite{av_indicate,d_av_gen,vid2speech,v2sound,audio_impaint,tip20_gen,eccv20_foley,eccv18_soundgen} have been proposed to utilize visual cues to generate audio outputs which match the sounding objects in the visual scene. For example, Owens~\etal~\cite{av_indicate} demonstrate that deep neural networks are capable of synthesizing new sounds for videos by looking at the material where the drumstick hits. In~\cite{vid2speech}, raw pixels of a speaker's face are mapped into audio features which are subsequently converted into an intelligible waveform. In addition, recurrent networks are also shown to be effective in generating audio input video frames~\cite{v2sound}. Generating audio data~\cite{avgen_cmcgan} can be realized by leveraging an encoder-decoder generative adversarial network (GAN)~\cite{goodfellow2014generative} conditioned on the visual frames. However, different from generating audio outputs associated with particular visual objects or scenes, the main focus of our method lies in converting single channel audio (i.e., monaural audio) into a dual channel one (i.e., binaural audio), which is guided by observing cross-modality features for improved performances.

\textbf{Audio-Visual Spatialization.} 
Audio-visual spatialization aims at separating the audio input to multiple outputs based on the locations of interest. Recently, a self-supervised neural network~\cite{360gen,iccv19_sptrack} is proposed to perform such tasks using videos with spatial audio recording. Given a 360$^\circ$ video with a single channel audio, their model learns to recover ambisonic audio outputs (i.e., four channels for the 360$^\circ$ video), enabling users to immerse sounds from all directions. To better capture the visual cues, their model exploits motion information for generating audio with better ambisonic quality. However, one cannot directly apply their approach for videos not recorded in the 360$^\circ$ format. 

In~\cite{25d}, Gao~\etal~propose a model that converts mono audio to stereo audio in 2D videos by measuring the difference between left and right audio channel outputs. This characteristic would guide the model to convert stereo audio of better quality. Moreover, Lu~\etal~\cite{25dclass} utilize not only visual and motion information but also includes a scene classifier which guides the generation of binaural audio with the associated scene label information. However, the use of their model would require additional scene label annotation during training, which may reduce the generalization of model (for unseen scenes, etc.). Although both~\cite{25d} and~\cite{25dclass} uitlize visual features in their model for predicting binaural audio outputs, their models are not designed to discover the spatial information of sounding regions corresponding to distinct audio components. Thus, their abilities for audio spatialization would still be limited (as confirmed later by our experiments).

\section{Method}
\label{sec:method}
\subsection{Problem Formulation and Notations}
We first define the notations and settings considered in this paper. 
As shown in Fig~\ref{fig:method}, the inputs of our network takes the monaural audio with a set of corresponding visual frames. Following the training setting in~\cite{25d,25dclass},
the monaural audio $x^M(t)$ input is mixed from binaural ground truth ones by adding left $x^L(t)$ and right sound $x^R(t)$ together at time $t$. To analyze such audio signals at distinct frequency bands, the input monaural audio is transformed into frequency domain by short-time Fourier transform (STFT)~\cite{stft}. That is, the mixed monaural $x^M(t)$ is transformed into the spectrogram $\mathbf{S}^M \in \mathbb{C}^{u \times t}$ (and later encoded as the audio feature $\mathbf{a}\footnote{note that $u,t$ is $2^n$ times smaller compared with spectrogram inputs for simplicity} \in \mathbb{R}^{d \times u \times t}$). As for the visual feature of each frame, it is represented by $\mathbf{v} \in \mathbb{R}^{d \times w \times h}$. Note that $d$ indicates the feature dimension for each channel; $u$ and $t$ indicate the size of the audio spectrogram, while $w$ and $h$ denote that of the visual feature. As for the outputs, our network produces left $\textbf{M}^{L}$ and right $\textbf{M}^{R}$ complex masks which decompose the recovered monaural sound spectrogram $\tilde{\textbf{S}}^{M}$ into left and right ones, respectively. In othe words, the predicted left $\tilde{\textbf{S}}^{L}$ and right $\tilde{\textbf{S}}^{R}$ complex spectrograms can be represented as: 
\begin{equation}
\begin{aligned}
\label{eq:pred_diff}
\tilde{\mathbf{S}}^{L} = \mathbf{M}^{L} \times \mathbf{S}^{M},\,\tilde{\mathbf{S}}^{R} = \mathbf{M}^{R} \times \mathbf{S}^{M}. 
\end{aligned}
\end{equation}

\subsection{Revisit of Spatial Audio Synthesizer}
Recently,~\cite{25d,25dclass} address audio spatialization and predict 
binaural audio outputs from videos with only monaural inputs. They adopt the U-Net~\cite{unet} which takes monaural audio as inputs and injects visual features extracted from videos at the bottleneck of U-Net, which guides the decoder to recover audio outputs with ground truth binaural audio observed. 

In stead of directly optimizing both binaural recording at left and right channels~\cite{25dclass}, Guo~\etal~\cite{25d} particularly train the network by measuring the difference of binaural recordings, in which the training objective is described as follows:
\begin{equation}
\label{eq:rec_loss}
\mathcal{L}_{rec}= {\lVert \tilde{\mathbf{S}}^{D} - \mathbf{S}^{D}\rVert}_2,
\end{equation}
\noindent where $\tilde{\mathbf{S}}^{D}=\tilde{\mathbf{S}}^{L} - \tilde{\mathbf{S}}^{R}$.
As for~\cite{25dclass}, additional information such as scene labels is taken into consideration for audio spatialization. However, collecting a large amount of video data with ground truth binaural audio and/or scene labels would be expensive. This is the reason why we choose to exploit information observed across spatial and audio modalities for better guiding the learning/prediction process.


\subsection{Exploiting Interaural Level Difference for Audio-Visual Consistency}
To convert monaural audio of a video into binaural ones, we propose to explore the correlation between particular audio components and visual regions across video frames for learning the audio spatialization model. This would not only guide the training of our model; more importantly, it would alleviate the need to collect a large number of videos with ground truth binaural audio for training. 

We note that, the characteristics of spatial audio (particular for binaural audio) relies on the difference between the audio signals received by left and right ears. To be more specific, the decisive factors of binaural recording originate from interaural time differences (ITDs) and interaural level differences (ILDs), which allow human beings to sense 3D surrounding audio in a scene. Therefore, it would be crucial in determining different levels of magnitude (energy) between the left-right channels of binaural recording and recovery. Comparing the audio signals received by the two channels, the channel with a larger magnitude indicates that the audio source is closer to the corresponding audio receiver. Such properties and observations also imply that the location of sounding objects can be possibly determined by measuring the correlation between audio frequency and visual patches associated with that object. 

Following the above idea, we first consider the difference between the audio signals (in magnitude) received by left and right channels. With the spectrogram signals predicted by the left and right channels, we calculate their difference by: 
\begin{equation}
\label{eq:mag_diff}
|\textbf{S}^{D}| =  |\textbf{S}^{L}| - |\textbf{S}^{R}|,
\end{equation}
where $|\textbf{S}^{D}|$ is of size $u \times t$, indicating the magnitude difference between the left-right spectrograms. If an entry in $|\textbf{S}^{D}|$ is greater than zero, it means that the sounding object with the corresponding frequency-time specific audio component is at the left-hand side of the scene. To normalize such difference values into probability values, we apply a sigmoid function as follows:

\begin{equation}
\label{eq:pl_gt}
\mathbf{P}^{L\rightarrow R}_{a} = Sigmoid\left( \dfrac {|\textbf{S}^{D}|}{\max(|\textbf{S}^{D}|)-\min(|\textbf{S}^{D}|)}\right).
\end{equation}
Note that $\mathbf{P}^{L\rightarrow R}_{a}$ is also of size $u \times t$, in which each entry indicates how likely the sounding source of the associated frequency-time specific audio component locates at the left hand side of the input scene. It is worth noting that, if ground truth $\textbf{S}^{L}$ and $\textbf{S}^{R}$ are not available during training, they will be replaced by the predicted ones $\tilde{\textbf{S}}^{L}$ and $\tilde{\textbf{S}}^{R}$ for guiding the training process as discussed later.

\subsection{Audio Spatialization with Audio-Visual Consistency}
\label{sec:co-att}
As pointed earlier, the key idea of our learning model lies in the observation of the spectrogram difference between each frequency-time specific audio component received by left and right ears. By jointly exploiting the correlation across audio spectrogram and visual features, the left-right location of sounding objects can be discovered accordingly.

To realize the above idea, we propose to learn the co-attention across audio and visual latent features, as depicted in Figs.~\ref{fig:method}. Given the $k$th patch in the audio spectrogram (out of $u \times t$ patches in $\mathbf{a} \in \mathbb{R}^{d \times (u \times t)}$), we calculate the correlation (in cosine similarity) between it and the visual feature $\mathbf{v} \in \mathbb{R}^{d \times (w \times h)}$ extracted from a video frame. As a result, the resulting co-attention map for $\mathbf{a}_k$ can be expressed as follows:
\begin{equation}
\label{eq:cos}
\begin{gathered}
   \mathbf{C}_{k} = \phi_{cos}(\mathbf{v}, \mathbf{a}_k), \forall k = \{1, ..., (u \times t)\},
\end{gathered}
\end{equation}
where $\mathbf{C} \in \mathbb{R}^{w \times h \times (u \times t)}$ indicate the correlation scores between $w \times h$ visual patches and $u \times t$ audio components in the spectrogram, and $\phi_{cos}$ denotes the cosine similarity function. 

Although the above co-attention map is derived between each audio component and the visual features, which only reflects the correlation between each monaural audio component and the associated visual frame. To further determine the left-right location information, we have each co-attention map multiplied by two sigmoid-like weighting functions $\mathbf{W}_L$ and $\mathbf{W}_R$, both in size $w \times h$ and each column sharing the same value. For example, $\mathbf{W}_R$ can be described as:
\begin{equation}
\begin{aligned}
\label{eq:sigmoid-w}
\mathbf{W}_{R}(:,x) = \frac{1}{1+e^{-\left(qx+r\right)}},
\end{aligned}
\end{equation}
where $q \in \mathbb{R}^{+}$ and $r \in \mathbb{R}$ are both constants. Similar remarks ($q \in \mathbb{R}^{-}$) can be applied for $\mathbf{W}_{L}$.


Multiplying the co-attention map by $\mathbf{W}_L$ and $\mathbf{W}_R$ would imitate the received signals of each sounding source (i.e., $\mathbf{a}_k$) reaching left and right channels. By measuring the difference between two scores (also normalized by sigmoid), the probability of each particular audio component locating at left (or right) hand side of a scene can be produced:  
\begin{equation}
\begin{gathered}
\label{eq:diff_vis}
p^{L\rightarrow R}_{a,v}(k) = Sigmoid \{ \max(\mathbf{W}_L\mathbf{C}_{k}) -\max(\mathbf{W}_R\mathbf{C}_{k}) \},  \\ \forall k = \{(1, ..., (u \times t)\},
\end{gathered}
\end{equation}
where $\mathbf{C}_{k} \in \mathbb{R}^{w \times h}$ is the correlation scores between audio patches and all $w \times h$ visual patches. $\mathbf{W}_L, \mathbf{W}_R \in \mathbb{R}^{w \times h}$ denote the functions weighting the co-attention maps. As a result, we would observe the left-right probability outputs for all audio frequencies as $\mathbf{P}^{L\rightarrow R}_{a,v} \in \mathbb{R}^{u \times t}$, based on the visual-audio correlation. Together with the left/right audio cues observed in~\eqref{eq:pl_gt}, we calculate the following loss function $\mathcal{L}_{con}$ for preserving the audio-visual consistency:
\begin{equation}
\begin{aligned}
\label{eq:loss_prob}
\mathcal{L}_{con} = BCE(\mathbf{P}^{L\rightarrow R}_{a,v}, \mathbf{P}^{L\rightarrow R}_{a}).
\end{aligned}
\end{equation}
Note that BCE represents the binary cross entropy calculation. As discussed earlier, if ground truth binaural audio outputs are not available during training, this consistency loss $\mathcal{L}_{con}$ can still be calculated, which can be viewed as a self-supervised learning technique. Nevertheless, if full supervision is available, our model can be trained by jointly observing the above $\mathcal{L}_{con}$ and the (ground truth) audio recovery loss $\mathcal{L}_{rec}$ as described in \eqref{eq:rec_loss}.

\section{Experimental Results}
\label{sec:results}

\begin{table*}[t]
\centering

\begin{tabular}{c|cc|cc|cc|cc}
\multirow{2}{*}{\diagbox[width=12em]{Method}{ Dataset }}   & \multicolumn{2}{c|}{FAIR-PLAY}         & \multicolumn{2}{c|}{REC-STREET}          & \multicolumn{2}{c|}{YT-CLEAN}          & \multicolumn{2}{c}{YT-MUSIC}  \\ \cline{2-9} 
                                & STFT             & ENV            & STFT             & ENV            & STFT             & ENV             & STFT             & ENV            \\ \hline
Mono                            & 1.155             & 0.153         & 0.774         & 0.136             & 1.369         & 0.153         & 1.853                 &0.184          \\
Audio Only                      & 0.966             & 0.141         & 0.590         & 0.114             & 1.065         & 0.131         & 1.553                 & 0.167          \\
Ambisonics~\cite{360gen}         & -              & -               & 0.744         & 0.126            & 1.435         & 0.155          & 1.885                 & 0.183          \\
Lu~\etal~\cite{25dclass}         & 0.899          & 0.139           & 0.568          & 0.109             & 1.032         & 0.130         & 1.459                 & 0.160          \\
MONO2BINAURAL~\cite{25d}         & 0.909          & 0.140           & 0.571          & 0.110             & 1.035         & 0.131         & 1.455                 & 0.162         \\ \hline \hline
Ours w/o $\mathcal{L}_{con}$     & 0.904          & 0.140           & 0.569          & 0.109             & 1.033         & 0.130         & 1.457                 & 0.161         \\
Ours                             & \textbf{0.865} & \textbf{0.136}  & \textbf{0.561} & \textbf{0.104} & \textbf{1.029} & \textbf{0.124} & \textbf{1.448} & \textbf{0.155}
\end{tabular}%
\caption{Performance comparisons of binaural audio prediction on four video datasets in terms of STFT and ENV. Note that the numbers in bold indicate the best results.
}

\label{tab:stoa}
\end{table*}

\subsection{Datasets}
\label{sec:dataset}
\textbf{FAIR-PLAY}~\cite{25d}. The FAIR-PLAY dataset consists of 1,871 10s clips of videos with binaural recording. These videos are recorded in a music room where reverberation has less influence in professional binaural recording. As for the train/val/test split, we follow up given splits from FAIR-PLAY dataset. 
\\
\textbf{REC-STREET}~\cite{360gen}. The REC-STREET dataset consists of 43 videos (3.5 hours) recorded in street scenes $360^\circ$ format with $1^{st}$ order ambisonic format audio (4 channels).
\\
\textbf{YT-CLEAN}~\cite{360gen}. The YT-CLEAN dataset contains 496 videos collected on YouTube in $360^\circ$ format both audio and visual content. The scenes of these videos vary such as meeting rooms, train carriages, restaurants, and etc.
\\
\textbf{YT-MUSIC }~\cite{360gen}. The YT-MUSIC dataset consists of 397 videos also collected on YouTube in $360^\circ$ format. Music and singing performance are recorded in these videos. The audio of videos is mixed with multiple similar sources like instruments and voices from different people. For $360^\circ$ videos, their audio encoding format is different from binaural recording as in 2D videos. Therefore, pre-processing for $360^\circ$ videos is required not only for comparison purposes but also for fitting the binaural setting. We follow~\cite{25d} and process the audio formats of $360^\circ$ videos. That is, the ambisonics (4 channels) recording is decoded into the binaural one using the transfer function (HRTF) from NH2 subject in the ARI HRTF Dataset2.

\subsection{Implementation Details} 
In all the experiments, only the visual frame corresponding to the middle of the audio segment is extracted. For example, the time of an audio segment is from 0.2 sec to 0.8 sec, and the visual frame to be considered is the one at time 0.4 sec. The visual feature is extracted from ResNet-18~\cite{resnet} which is pre-trained on ImageNet~\cite{ImageNet}.

We implement our model using PyTorch~\cite{pytorch} and train our model on a single NVIDIA GTX 1080 Ti GPU with $12$ GB memory. To fairly compare with the baseline methods, our model utilizes the same number of model parameters. The performance of our method can be possibly further improved by adding more layers for U-Net based spatial audio synthesizer or replacing with different U-Net backbone. However, such techniques are not used in all of our experiments.

As for audio settings in our experiments, the raw audio data are resampled at 16kHZ. As for the STFT setting, we use a Hann window of length 25ms, FFT size of 512 and hop length of 10ms. During training, we randomly sample one audio segment with 0.63s in a video with the corresponding video frame. As for testing, we sample all the audio segments in a video with 0.05s hop size. 

\subsection{Evaluation Metrics}
As considered in~\cite{25d,25dclass,360gen}, two evaluation metrics are utilized for measuring the recovered spatial audio quality.
\\
\textbf{STFT Distance: }We computes the Euclidean distance between the ground-truth complex spectrograms and predicted one which are scaled back as raw audio energy level. The left and right are both evaluated:
\begin{equation}
\label{eq:stftl2}
\mathcal{D}_{STFT}= {\lVert \tilde{\mathbf{X}}^{L} - \mathbf{X}^{L}\rVert}_2 + {\lVert \tilde{\mathbf{X}}^{R} - \mathbf{X}^{R}\rVert}_2.
\end{equation}
\textbf{Envelope (ENV) Distance: }
In time domain, directly measuring raw waveforms may not capture perceptual similarity well. We compute the envelope of ground-truth and predicted waveform, and measure their the Euclidean distance: 
\begin{equation}
\begin{aligned}
\label{eq:envl2}
\mathcal{D}_{ENV}&= {\lVert \mathrm{E}(\tilde{\mathbf{x}}^{L}(t)) - \mathrm{E}(\mathbf{x}^{L}(t))\rVert}_2 \\&+ {\lVert \mathrm{E}(\tilde{\mathbf{x}}^{R}(t)) - \mathrm{E}(\mathbf{x}^{R}(t))\rVert}_2,
\end{aligned}
\end{equation}
where $\mathrm{E}(.)$ denotes the envelope of signal $x(t)$.

\captionsetup[subfigure]{labelformat=empty}
\begin{figure*}[t!]
\centering
  \begin{subfigure}[b]{.24\linewidth}
    \centering
    \includegraphics[width=.99\textwidth]{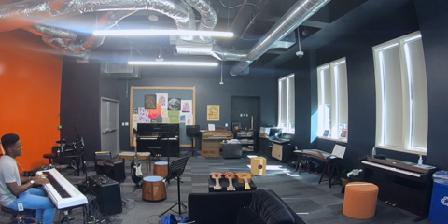}
  \end{subfigure}%
  \begin{subfigure}[b]{.24\linewidth}
    \centering
    \includegraphics[width=.99\textwidth]{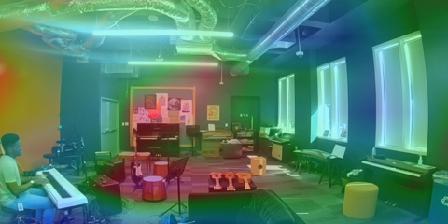}
  \end{subfigure}%
  \begin{subfigure}[b]{.24\linewidth}
    \centering
    \includegraphics[width=.99\textwidth]{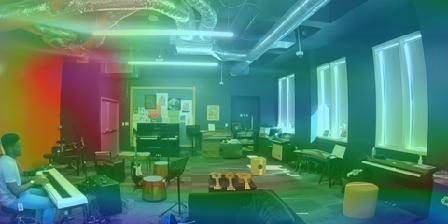}
  \end{subfigure}%
  \begin{subfigure}[b]{.24\linewidth}
    \centering
    \includegraphics[width=.99\textwidth]{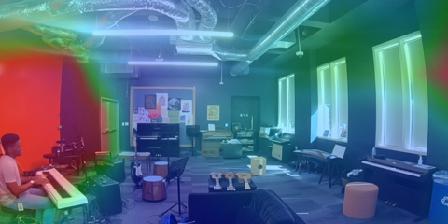}
  \end{subfigure}\\%
    \begin{subfigure}[b]{.24\linewidth}
    \centering
    \includegraphics[width=.99\textwidth]{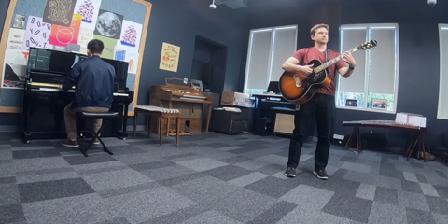}
  \end{subfigure}%
  \begin{subfigure}[b]{.24\linewidth}
    \centering
    \includegraphics[width=.99\textwidth]{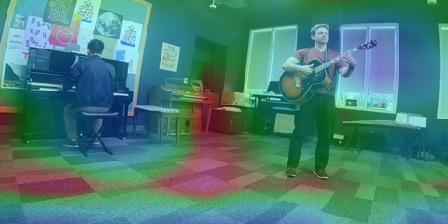}
  \end{subfigure}%
  \begin{subfigure}[b]{.24\linewidth}
    \centering
    \includegraphics[width=.99\textwidth]{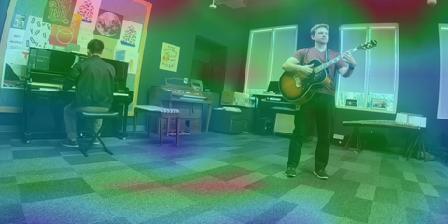}
  \end{subfigure}%
  \begin{subfigure}[b]{.24\linewidth}
    \centering
    \includegraphics[width=.99\textwidth]{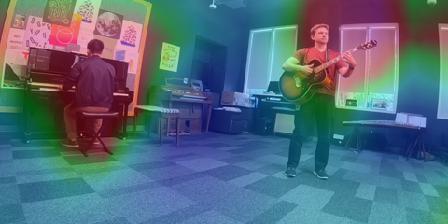}
  \end{subfigure}\\%
  
  \begin{subfigure}[b]{.24\linewidth}
    \centering
    \includegraphics[width=.99\textwidth]{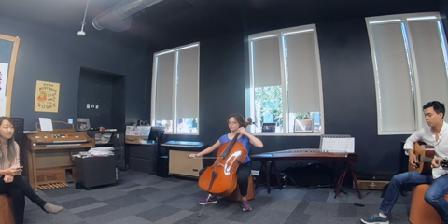}
    \caption{Input Frame}
  \end{subfigure}%
  \begin{subfigure}[b]{.24\linewidth}
    \centering
    \includegraphics[width=.99\textwidth]{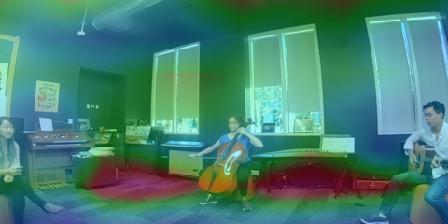}
    \caption{MONO2BINAURAL}
  \end{subfigure}%
  \begin{subfigure}[b]{.24\linewidth}
    \centering
    \includegraphics[width=.99\textwidth]{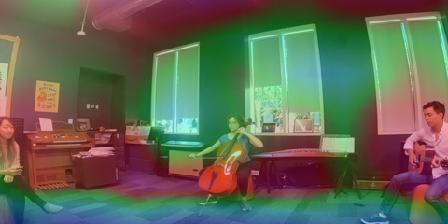}
    \caption{Lu~\etal}
  \end{subfigure}%
  \begin{subfigure}[b]{.24\linewidth}
    \centering
    \includegraphics[width=.99\textwidth]{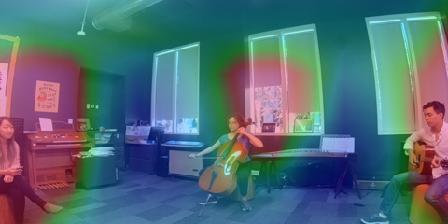}
    \caption{Ours}
  \end{subfigure}\\%
  \caption{Example visualization results on FAIR-PLAY. Note that the audio-visual attention is depicted in terms of heat maps, in which red regions indicate high correlation between audio and visual data. More visualization results including demo videos are available in the supplementary. }\label{fig:vis}
\end{figure*}

\subsection{Quantitative Evaluation}
To evaluate the quality of our predicted binaural audio, we compare our model with the following baselines or state-of-the-arts methods:
 \begin{itemize}
\item\textbf{Audio Only:} The model is trained without any visual frame information, which directly predicts binaural audio outputs given mixed monaural audio only.

\item \textbf{Mono:} The mixed monaural audio is directly replicated onto the left and right audio channel to create fake binaural audio which preserves no spatial information.

\item \textbf{Ambisonics~\cite{360gen}:} As the state-of-the-art method for the B-format (4 channels for $360\circ$ videos) audio generation, we first reproduce spatial audio in B-format with the pre-trained models. Then, the reconstructed spatial audio is decoded into binaural format by the HRTF decoder. Owing to the limitation of generation binaural recording, it cannot be applied to the FAIR-PLAY dataset.

\item \textbf{Lu~\etal~\cite{25dclass}:} This approach jointly considers audio, visual and flow information extracted by FlowNet~\cite{flownet2}, plus a scene classifier providing additional guidance. Since no scene annotation is available for the datasets considered in this paper, we simply remove the scene classifier in our experiments during performance comparisons (note that all video scenes in FAIR-PLAY are the same).  

\item \textbf{MONO2BINAURAL~\cite{25d}:} Also considered as a state-of-the-art method for spatial audio generation, which considers audio-visual information at the bottleneck of their model and requires full supervised during training.

\end{itemize}

Table~\ref{tab:stoa} compares our model with the aforementioned baseline and three state-of-the-art methods for binaural audio prediction. From this table, it is clear that our method performed favorably against the state-of-the-art approaches on all four datasets, including the ones utilizing visual information. This supports our exploitation of audio-visual co-attention for guiding the learning of particular audio components (at left-right channels) with the associated location in a scene. We note that, however, the performance improvement on the $360^\circ$ videos (REC-STREET, YT-CLEAN and YT-MUSIC) was marginal. The main reasons are as follows: the datasets are not real binaural recording (somehow simulated) which is decoded by HRTFs that would lose some details because HRTFs simulate position of human beings in sphere space. Furthermore, there are multiple sources with silent audio segments throughout the videos, which eventually increase the difficulty of generating spatial audio. Nevertheless, our model still achieved satisfactory performances when comparing to the state-of-the-art ones.

It is worth noting the effect of optical flow estimation when viewing videos and generating the binaural audio outputs. Since the scene label is not available, the main difference between the works of~\cite{25dclass} and~\cite{25d} would be the use of flow information. Based on the experimental results, we only observe the exploitation of flow information to be marginal in performing this task. We believe that the possible reason would be the audio delivered by the sounding objects might not be highly correlated with their movements, and thus utilizing flow information would not be sufficiently meaningful. And, due to space limitation, more quantitative results will be provided in the supplementary materials.

\subsection{Qualitative Evaluation} 


The visualization results are shown in Fig.~\ref{fig:vis} on videos selected from the FAIR-PLAY dataset. From this figure, we see that our model better associates audio data with the sounding objects. The example in the bottom row is particularly challenging, since there are multiple sounding objects, which makes the exploitation of correlation between visual and audio data more difficult. Nevertheless, from the examples shown in this figure, we see that our model was able to identify the sounding regions of interest when comparing to state-of-the-art methods. It is worth noting that, no ground truth spatial information is available for all sounding objects. Thus, these results support the use of our model for discovering the sounding sources in a scene, which would be applied for identifying audio signals received by left and right channels as described in audio spatialization with audio-visual consistency section.



\begin{table}[]

\centering
\begin{tabular}{c|cccc|cc}
\multicolumn{1}{l|}{Layers}                                                               & 1          & 2          & 3          & 4          & STFT                      & ENV                        \\ \hline
\multirow{7}{*}{\rotatebox[origin=c]{90}{Utilized Layers}} & \checkmark &            &            &            & 0.865                     & 0.1363                     \\
                                                                                          &            & \checkmark &            &            & 0.868                     & 0.1365                     \\
                                                                                          &            &            & \checkmark &            & 0.872                     & 0.1371                     \\
                                                                                          &            &            &            & \checkmark & 0.878                     & 0.1375                     \\ \cline{2-7} 
                                                                                          & \checkmark & \checkmark &            &            & \multicolumn{1}{l}{0.868} & \multicolumn{1}{l}{0.1366} \\
                                                                                          & \checkmark & \checkmark & \checkmark &            & \multicolumn{1}{l}{0.871} & \multicolumn{1}{l}{0.1369} \\
                                                                                          & \checkmark & \checkmark & \checkmark & \checkmark & \multicolumn{1}{l}{0.869} & \multicolumn{1}{l}{0.1367}
\end{tabular}
\caption{Performance comparisons on FAIR-PLAY using audio features extracted at different decoder layers from the U-Net like auto-encoder architecture in Fig.~\ref{fig:method}. Note that check mark indicates layers which are utilized.}
\label{tab:abs}
\end{table}

\begin{figure*}[t!]
	\centering
	\begin{subfigure}[h]{0.45\linewidth}
    \includegraphics[width=\linewidth]{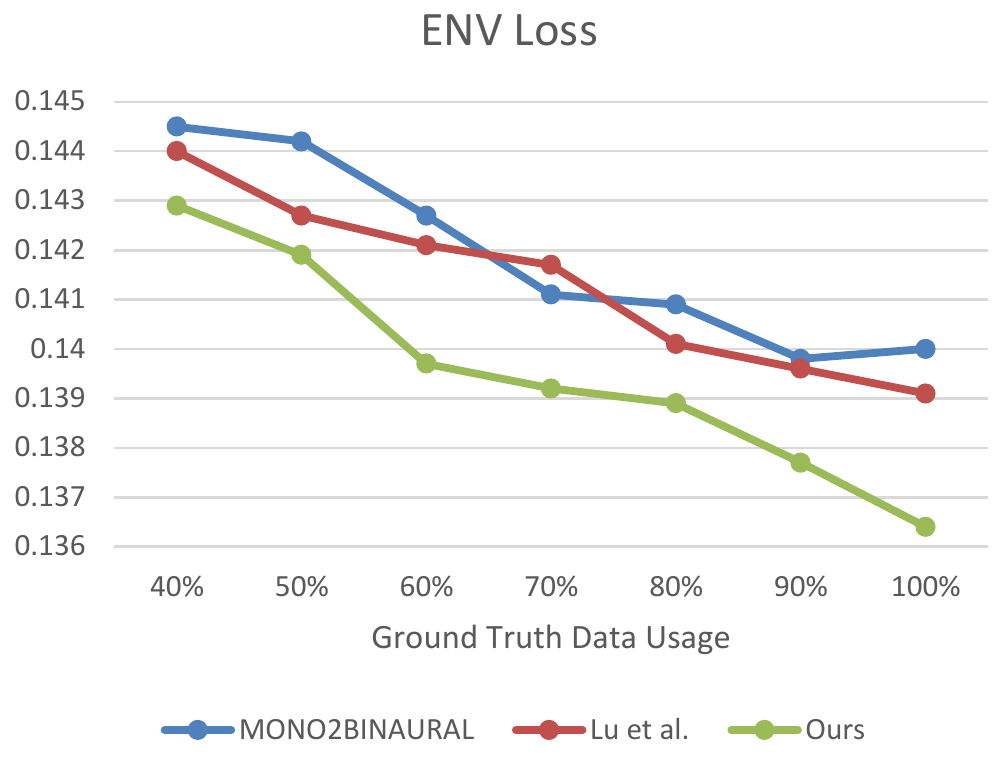}
    \caption{Envelope (ENV) Distance}
    \label{fig:semi_env}
    \end{subfigure}
    \hfill
    \begin{subfigure}[h]{0.45\linewidth}
    \includegraphics[width=\linewidth]{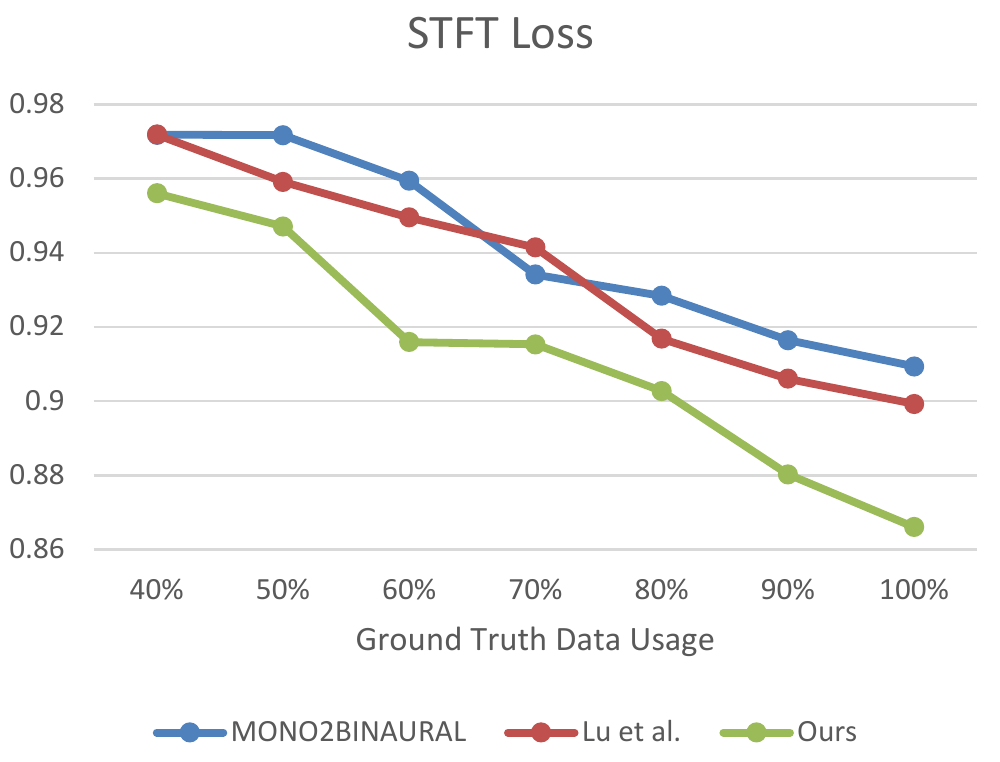}
    \caption{STFT Distance}
    \label{fig:semi_stft}
    \end{subfigure}%
     \caption{Performance comparisons in terms of ENV and STFT on FAIR-PLAY using different amount of video data with ground truth binaural audio for training. Note that the horizontal axes in both figures reflects the percentage of data with ground truth audio utilized. It can be seen that our model achieved comparable results as other fully supervised models did, while only about 60\% of labeled data were used when training our model.}
	\label{fig:semi}
\end{figure*}
\subsection{Ablation Study}
Since our model utilizes the proposed audio-visual consistency loss~\eqref{eq:loss_prob} on state-of-the-arts models like~\cite{25d} for spatial audio generation, Table~\ref{tab:stoa} already compares and verifies the contribution of this proposed loss term. On the other hand, since such observed consistency comes from the co-attention of audio-visual features, we now evaluate the performance of our model using audio features extracted from different layers (for~\eqref{eq:loss_prob}) in the decoder of our U-Net like architecture. Table~\ref{tab:abs} lists both STFT and ENV results using audio features at various layers. From this table, we see that the use of the latent feature extracted at the highest-level single layer would be preferable, while using multiple cross-scale audio features were not able to achieve comparable results. This is due to the fact that the U-Net architecture outputs binaural outputs, and thus extracting features at finer layers would not contain spatial specific features, which would be redundant for calculating audio-visual co-attention for observing the left-right consistency in~\eqref{eq:loss_prob}.

\subsection{From Supervised to Semi-Supervised Learning}

As discussed earlier, since our model is trained by jointly minimizing binaural audio recovery and spatial-audio consistency losses, our model can be realized in semi-supervised settings. That is, only a portion of video data is with ground truth binaural audio while the remaining ones are unlabeled. It is worth repeating that, the ground truth spatial information of sounding objects is never observed during training. As a result, we choose to vary the percentage of ground truth binaural recording for training, and present the results in Fig.~\ref{fig:semi}. From the results shown in this figure, we see that the use of our spatial-audio consistency loss (i.e., the exploitation of spatial-audio co-attention) would benefit binaural audio prediction. When the amount of labeled audio data increases, all methods especially ours would better learn the relationship between visual and audio information presented in a video. The gap between our method with~\cite{25dclass} and~\cite{25d} would be appreciable from this figure. For example, using only 60\% of labeled data, our model was able to achieve comparable performances as~\cite{25dclass} and~\cite{25d} did. Therefore, from the above experiments, the use of our model for binaural audio prediction in a semi-supervised setting can be successfully verified.

\section{Conclusions}
\label{sec:conclusions}
In this paper, we presented a novel framework to generate binaural audio from the input video with only monaural recording. The novelty of our proposed model lies in the ability in exploiting the correlation between each audio component and the spatial regions of interest, which would guide the learning of left-right audio difference during training. Since no ground truth spatial information is observed in the above process, our learning scheme can be viewed as a self-supervised learning technique, and thus can be integrated to existing binaural audio recovery models (with full supervision of ground truth binaural audio). Moreover, our learning strategy further alleviates the dependency of models learned in fully supervised settings, and thus can be realized in semi-supervised settings with promising performances. Our experimental results quantitatively and qualitatively support the use of our model, confirming its superiority over state-of-the-arts models in both supervised and semi-supervised settings.

\section{Acknowledgements}
This work is supported in part by the Ministry of Science and Technology of Taiwan under grant MOST 109-2634-F-002-037.
\bibliography{ref}
\end{document}